\documentclass[conference]{IEEEtran}
\IEEEoverridecommandlockouts
\usepackage{cite}
\usepackage{amsmath,amssymb,amsfonts}
\usepackage{algorithmic}
\usepackage{graphicx}
\usepackage{textcomp}
\usepackage{xcolor}
\usepackage{multirow}
\usepackage{makecell}
\usepackage[hidelinks]{hyperref}
\usepackage{float}
\usepackage{enumitem}

\usepackage[skip=5pt,belowskip=-2pt]{caption}
\def\BibTeX{{\rm B\kern-.05em{\sc i\kern-.025em b}\kern-.08em
    T\kern-.1667em\lower.7ex\hbox{E}\kern-.125emX}}
\begin{document}

\title{Development of Reduced Feeder and Load Models Using Practical Topological and Loading Data  
}

\author{
\IEEEauthorblockN{Sameer Nekkalapu\textsuperscript{1}, Sushrut Thakar\textsuperscript{2}, Antos Cheeramban Varghese\textsuperscript{3}, Vijay Vittal\textsuperscript{3},  Bo Gong\textsuperscript{4}, Ken Brown\textsuperscript{4} \\
\textsuperscript{1}\textit{Pacific Northwest National Laboratory, WA, USA};  \textsuperscript{2}\textit{Electric Power Research Institute, TN, USA}; \\  \textsuperscript{3}\textit{Arizona State University, AZ, USA}; \textsuperscript{4}\textit{Salt River Project, AZ, USA}} 
\and
\thanks{This work has been funded by Salt River Project.}
}
\maketitle

\begin{abstract}
Distribution feeder and load model reduction methods are essential for maintaining a good tradeoff between accurate representation of grid behavior and reduced computational complexity in power system studies. An effective algorithm to obtain a reduced order representation of the practical feeders using utility topological and loading data has been presented in this paper. Simulations conducted in this work show that the reduced feeder and load model of a utility feeder, obtained using the proposed method, can accurately capture contactor and motor stalling behaviors for critical events such as fault induced delayed voltage recovery. 
\end{abstract}

\begin{IEEEkeywords}
Contactor, Distribution, Loads, Modeling, Motors, Reduced Feeder, Utility Data  
\end{IEEEkeywords}

\section{Introduction}

Representation of the distribution system behavior in power system studies requires accurate feeder and load models \cite{price1993load_etal}.  The composite load model (CMPLDWG) is the state-of-the-art model for incorporating loads adequately into system planning and operation studies 
\cite{NERC_DLM_Report}. 
However, the composite load model represents the distribution feeders as an aggregated load at a single bus behind a feeder impedance \cite{nekkalapu2022emtp}. Currently, efforts are being undertaken by the industry vendors to modularize the composite load model to improve the model flexibility in terms of better representation of the load behavior such as the motor stalling phenomenon which leads to fault induced delayed voltage recovery (FIDVR) type critical events in the system \cite{liu2013transient_etal}. This is because the current version of the composite load model do not capture partial motor stalling behavior in which typically only motors in some sections of the feeder stalls depending on the voltage magnitude and the angle seen at their terminals at the moment of fault initiation \cite{liu2013transient_etal}, \cite{nekkalapu2023development_etal}. It is especially important to capture this partial motor stalling phenomenon in the presence of their protection models, contactors \cite{nekkalapu2022emtp}, to not overestimate or underestimate the resiliency of the distribution system under critical events such as the FIDVR type phenomenon.

    Therefore, in recent literature, a state-of-the-art three-segment feeder and load model has been developed for modeling the distribution systems behavior, in the grid \cite{nekkalapu2021synthesis, nekkalapu2023development_etal, nekkalapu2022emtp}. This feeder and load model has the capability to accurately capture both the necessary phenomenon – partial motor stalling and accurate contactor behavior, to accurately estimate the feeder behavior for FIDVR type critical events. However, this is dependent on having an accurate estimate of the feeder and load parameters for this model. In \cite{nekkalapu2021synthesis}, an optimization algorithm has been proposed to estimate the load parameters provided in this paper and in \cite{nekkalapu2023development_etal}, a data-driven deep neural network technique has been developed to estimate the contactor and motor stalling parameters. However, all these parameters were obtained by making simplistic assumptions about the voltage drops across the three-segment feeder model without validating it using field topological or measurement data across the feeders. Therefore, the main objective of this paper is to develop a systematic approach, while utilizing practical field data, to obtain a realistic reduced three-segment feeder model that can accurately capture the dynamic characteristics of a feeder comprising of contactors and the motor load behavior.

  The authors of \cite{wang2018time_etal} model a real network including 15 feeders in detail and study the FIDVR response and motor stalling.  However, they do not conduct dynamic simulations but rather rely on time series power flow to estimate the FIDVR response. The increased network size by modeling all the distribution feeders in detail (including the entire network) may not be suitable for dynamical simulations of larger networks owing to computational complexity.

	This computational complexity issue can be mitigated by reducing the distribution system model to an equivalent lower order model. \cite{casolino2020reduced} proposed a load area-based approach for the simplification of the distribution system. Generalized analytical methods were proposed in \cite{ pecenak2017multiphase_etal} to create a reduced order feeder model. However, \cite{casolino2020reduced, nowak2019measurement_etal, pecenak2017multiphase_etal} do not capture the dynamic behaviors of the distribution systems. A ruin and reconstruct-based feeder reduction method is proposed in \cite{nagarajan2017network_etal} for modeling feeders in real-time simulators. Although, these methods are good ways to reduce distribution system model order, they have not been tested using field data. 
    
    None of these approaches\cite{thakar2022mitigation_etal, wang2018time_etal, casolino2020reduced, nowak2019measurement_etal,  pecenak2017multiphase_etal} considered using utility topological data to obtain an accurate feeder structure which would be crucial in the process of the load modeling studies.
	This paper proposes an effective method to model a three-segment feeder model for distribution load modeling studies using practical topological feeder data obtained from a local utility in the South Western U.S. The proposed reduced feeder model is extremely useful for applications such as load composition and load parameter estimation \cite{nekkalapu2022improved} and contactor modeling \cite{nekkalapu2023development_etal}. 
    
The key contributions of this paper are as follows:

\noindent(1) An algorithm based on utilizing the feeder topological and loading data to reduce the distribution system into three segments is proposed. This reduced-order feeder model can be leveraged by the utilities to conduct accurate distribution system studies without any computational complexity issues that come with representing detailed distribution feeder models.\\ 
(2) The impact of utilizing the proposed modified three-segment feeder model to estimate the contactor behavior and motor stalling phenomenon accurately on the feeder behavior under critical FIDVR type phenomena has been demonstrated.

In this paper, Section \ref{Section_contactor_modeling_background}  provides the critical parameter details of the contactors and motor models used in this work. In Section \ref{Section_proposed_Approach}, the proposed methodology to create the reduced feeder model and its validation using the topological and loading data obtained from the local utility has been presented. Section \ref{Section_Simu_and_Results} investigates the performance comparison between the proposed modified three-segment feeder model with the existing state of the art three-segment feeder model \cite{nekkalapu2023development_etal} in the literature. The conclusions of this work are present in Section \ref{Section_conclusions}. All the simulations in this work have been conducted in an electromagnetic transient (EMT) simulator (PSCAD)  \cite{manitoba1994pscad}.
\section{Case Setup: Contactor \& SPHIMS Stalling Modeling}
\label{Section_contactor_modeling_background}
The three-segment three-phase feeder and load model considered in this work has been presented in Fig. \ref{Feeder_Structure}. The considered feeder model comprises of three-phase motor load, single-phase motor load (SPHIMs), resistive load and distribution transformers that step down the voltage from feeder head level (12.47 kV) to their respective rated low voltage load levels. More description about the contactor and the SPHIMs models modeling has been presented below -

\subsection{ Contactor Modeling}
A Contactor is an electromechanical device which trips the load (SPHIMs in this paper) under low voltage conditions. Typically, when there is a fault in the system, the voltage at the terminals of the contactor (and load) is depressed which leads to the contactor tripping or chattering. The chattering of the contactor refers to the phenomenon where the contactor repeatedly trips and reconnects when the voltage at its terminals is not low enough to trip it completely. 
In \cite{nekkalapu2022emtp}, an analytical model of an EMT contactor model has been developed. 

In this work, based on \cite{nekkalapu2022emtp}, \cite{nekkalapu2023development_etal}, the most important behavioral characteristics of a typical contactor model have been chosen and has been categorized as described below –

(i) Status of the contactor (‘\textit{ST}’): Under low voltage conditions, a contactor model either trips, chatters or remain unaffected.
 
(ii) Tripping and Reconnection Characteristics: The moment at which the contactor trips the load when the voltage at the terminals of the load goes below a certain threshold, and the moment at which the contactor reconnects the tripped load when the voltage at the terminals of the load recovers above a given threshold, corresponds to the tripping and reconnection characteristics respectively. Based on these characteristics, the following contactor features are identified and are described below – 

\begin{itemize}[leftmargin=2pt]
\item Tripping Parameters (‘\textit{T1}’ and '\textit{V1}’): These characteristics correspond to the time taken and the feeder head voltage level for the contactor to electrically disconnect the load or starts to chatter after the fault is applied in the system respectively.
\item Reconnection Parameters (‘\textit{T2}’ and '\textit{V2}’): These characteristics correspond to the time taken and the feeder head voltage level for the contactor to electrically reconnect the load after the fault is applied in the system respectively.
\end{itemize}


A pictorial description of the above trip and reconnection settings of the proposed contactor model is presented in Fig. \ref{Fig1_Contactor_setting}, for a scenario in which a single-line to ground (SLG) fault has been placed at the 69 kV sub-transmission level, as a function of the feeder head voltage (in blue), where all the contactors along the feeder have tripped (represented by the signals, in red, of the contactors changing from 0 to 1 in Fig. \ref{Fig1_Contactor_setting}) along the feeder on the faulted phase. In Fig. \ref{Fig1_Contactor_setting}, it should be noted that only the SPHIMs on phase A (faulted phase) have stalled in all three segments (and all three contactors on phase A tripped completely) and the SPHIMs in the remaining phases are unaffected for the considered SLG fault. 

\subsection{SPHIMS Stalling}
It has been widely studied in the literature that during some fault conditions, residential air conditioner SPHIMs stalling phenomenon occurs, in which the stalled motors speeds goes to zero due to their low inertia and draws enormous amount of reactive power in locked rotor condition. This leads to further worsening of the voltage in the system leading to a delayed voltage recovery (FIDVR phenomenon) even after the fault has been cleared. Using \cite{nekkalapu2023development_etal}, the stall parameters for the SPHIMs in this work has been defined as \textit{TMS} and \textit{IMS}. Where, \textit{TMS} corresponds to the total number of segments along the feeder in which the SPHIMs stall and \textit{IMS} is a vector containing information about the three individual segments of the feeder in which either the SPHIMs stall or do not stall. The possible values of \textit{TMS} can be \{0,1,2,3\}. Whereas, the possible values of \textit{IMS} can be \{\{0,1\},\{0,1\},\{0,1\}\}, in which \{0,1\}  corresponds to SPHIMs either stalling (1) or not stalling (0) respectively.   

\begin{figure}
    \centering
    \includegraphics[width=0.5\textwidth]{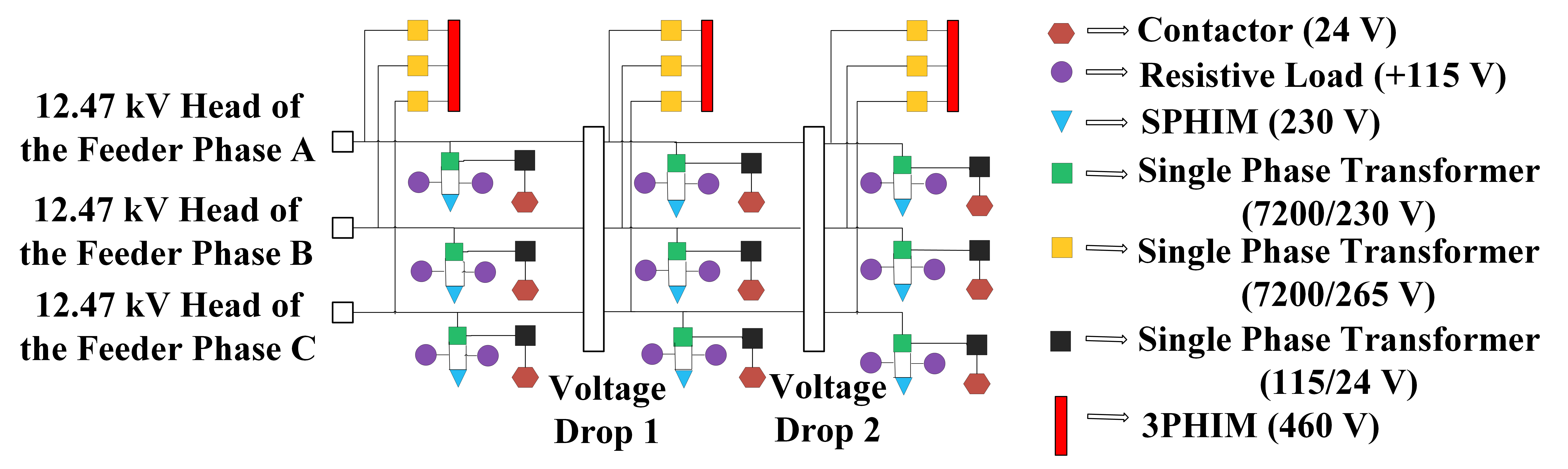}
    \caption{\textcolor{black}{Three segment feeder and load model structure considered in this work }} 
    \label{Feeder_Structure}
    \vspace{-1em}
\end{figure}

\begin{figure}[t]
    \centering
    \includegraphics[width=0.40\textwidth]{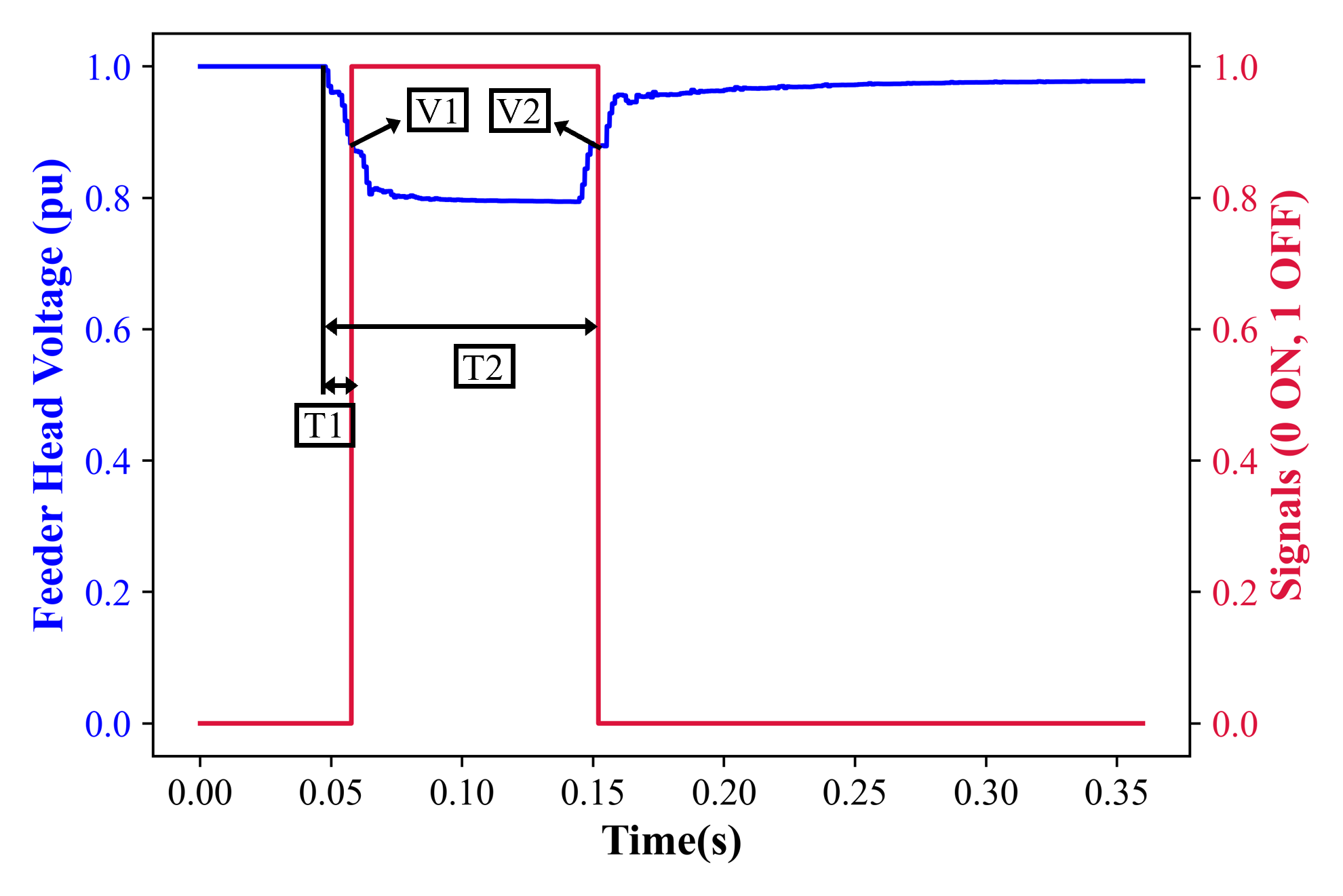} 
    \caption{\textcolor{black}{Demonstration of Contactor Settings to be Estimated as a Function of the Positive Sequence Feeder Head Voltage}} 
    \label{Fig1_Contactor_setting}
    \vspace{-1em}
\end{figure}

\section{Procedure for the Proposed Algorithm}
\label{Section_proposed_Approach}
It is important to validate \textcolor{black}{the critical features data of  contactors and motor stalling} (generated in PSCAD) for a three-phase three-segment feeder and load model considered in this work \cite{nekkalapu2022emtp}, \cite{nekkalapu2023development_etal}, \cite{nekkalapu2021synthesis}. This validation can been done using the topological details and the loading conditions obtained from a practical feeder. Therefore, in collaboration with a local utility in the South-Western U.S\textcolor{black}{.}, the practical feeder information \textcolor{black}{from a substation (\textcolor{black}{``}Substation A\textcolor{black}{''}) }in a residential area for highly loaded summer condition has been utilized in this work. It was observed that there are four three-phase 12.47 kV feeders on the low voltage side of Substation A. \textcolor{black}{The following information has been utilized to develop the proposed algorithm}:

\begin{enumerate} 
\item	Total Feeder Head MVA: It is important to know the total MVA drawn by all the distribution feeders originating from substation A. The aggregated MVA \textcolor{black}{values are} used in the proposed  model to simulate the \textcolor{black}{\textit{pre-fault steady state conditions}}. 
\item	Three-Section Segmentation: To mimic the proposed three-phase three-segment model in this work, the loads across the feeders need to be segmented into three sections appropriately based on their distribution \textcolor{black}{along the feeders. This is done by identifying the "load pockets" along the feeder where typically major portions of load would be concentrated at. This is especially important to accurately estimate the \textit{amount} of SPHIMs that would stall for a particular type of FIDVR event.} 


\item	Voltage Drop Data: Based on \textcolor{black}{this} three-part segmentation, the voltage drops across the three sections for the feeders need to be obtained \textcolor{black}{accurately to determine \textit{if} the SPHIMs across the three-segments would stall or not}.
\end{enumerate}

For the sake of simplification of the analysis, the above-mentioned criteria \textcolor{black}{have} been applied to the topology of the largest feeder (feeder A) among the four feeders from Substation A.\textcolor{black}{ The proposed algorithm has been described in detail  in the steps presented below}:

\subsection{Step 1: Nodes \& Sectional Data }
\textcolor{black}{The nodal topological data of the considered practical feeder have been categorized into various sections and their corresponding \textcolor{black}{`\textit{From}' and `\textit{To}'} nodes (buses) along the feeder}. \textcolor{black}{Here, the `\textit{From}' and `\textit{To}' nodes represent the direction of power flow through the section.} Although, in total, there are 485 sections and correspondingly 478 different nodes in feeder A, only 90 sections (180 nodes) have loads present on them. Therefore, it is important to \textcolor{black}{understand} \textcolor{black}{relative locations and distribution of these} loads (load pockets) along the feeder \textcolor{black}{as well as the} general topology \textcolor{black}{of the feeder.} Good visual representation of the feeder is necessary \textcolor{black}{for understanding these feeder characteristics}.


\subsection{Step 2: Feeder visualization}
To visualize feeder A, all the node data for feeder A were converted into a sparse symmetrical matrix form (478x478 dimensions) where all the columns and rows spanning the matrix correspond to \textcolor{black}{‘\textit{From}' and ‘\textit{To}’ nodes} respectively. Here, the matrix elements $m_{ab}$ are populated using \eqref{Matrix_Data}.

\begin{equation}
m_{ab} = 
\begin{cases}
1, & \text{if } n_{ab} \text{ exists}, \\
0, & \text{if } n_{ab} \text{ \textcolor{black}{does} not exist}.
\end{cases}
\label{Matrix_Data}
\end{equation}
where, $n_{ab}$ is the section between nodes $a$ and $b$.

\begin{figure}[h]
    \centering
    \includegraphics[width=0.485\textwidth]{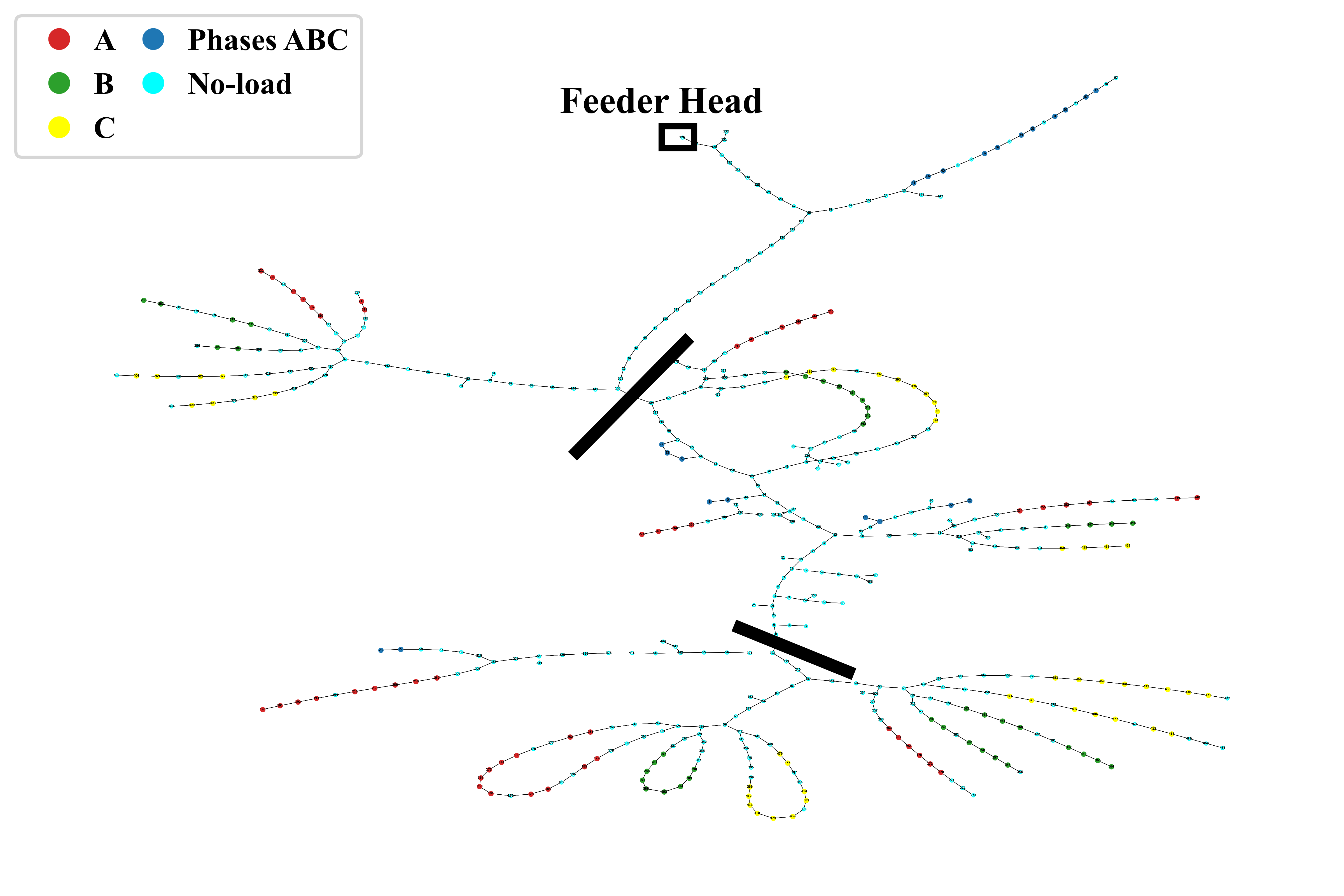}
    \caption{\textcolor{black}{Distribution feeder visualization}} 
    \label{Fig2_DF_Viz}
     \vspace{-0.5em}
\end{figure}

The populated sparse matrix is then read through a python script which uses directed graphs to \textcolor{black}{visualize} the feeder topology. ‘Kamada-Kawai’ \cite{kamada1989algorithm} force-directed layout library function was used in this work to generate the \textcolor{black}{ feeder A} graph, presented in Fig. \ref{Fig2_DF_Viz}. It should be noted that feeder A has various lateral feeders (which supply power to a small area) along its path. \textcolor{black}{Various types of lateral feeders considered in this work has been categorized}, connected to each colored node, into Cyan (no lateral feeders), Red (phase A lateral feeders), Green (phase B lateral feeders), Yellow (phase C lateral feeders) and Blue (three-phase lateral feeders).  


In Fig. \ref{Fig2_DF_Viz}, the head of the feeder node for feeder A has been marked with a rectangle. \textcolor{black}{Additionally,} based on the identified load pockets, the feeder has been divided into three parts using two solid black lines.


\begin{table}[]
\centering
\caption{Sequence Impedance Data for the Conductors along the Feeder}
\label{Table2_}
\begin{tabular}{|c|c|c|}
\hline
\begin{tabular}[c]{@{}c@{}}Conductor   \\      Type\end{tabular} & \begin{tabular}[c]{@{}c@{}}\textit{R1c} \\      (ohms/mile)\end{tabular} & \begin{tabular}[c]{@{}c@{}}\textit{X1c} \\      (ohms/mile)\end{tabular} \\ \hline
Type A                                                            & 1.91                                                            & 0.37                                                            \\ \hline
Type B                                                       & 0.63                                                            & 0.38                                                            \\ \hline
Type C                                                        & 0.25                                                            & 0.21                                                            \\ \hline
Type D                                                       & 0.23                                                            & 0.31                                                            \\ \hline
Busbar                                                           & 0                                                               & 0                                                               \\ \hline
\end{tabular}
\vspace*{-1\baselineskip}
\end{table}

\subsection{Step 3: Voltage Drop Calculations}

\textcolor{black}{Finally,} the voltage drop calculations across the feeder \textcolor{black}{are calculated} based on the topological data from Step 1 \& 2, and sectional impedance data
as shown in Table \ref{Table2_} for the four types of conductors and Busbars present in the feeder. Additionally, a feeder head voltage of 1.02 pu was assumed 
\textcolor{black}{because typically head of the  distribution feeders are operated at slightly higher nominal voltages to ensure that the loads at the end of the feeder do not see a voltage below 0.95 pu even during summer heavy loading conditions \cite{PSERC1}. Therefore, a three-phase capacitor bank has been placed at the feeder head to regulate the reactive power and maintain this feeder head voltage at steady-state conditions and a total of 3.63 MVA was observed to be drawn at the head of the feeder}. Using Kirchoff current and voltage laws along the feeder sections, the current flows and the voltage drops along the feeder have been calculated using only positive sequence impedance data (from Table \ref{Table2_}) and the active \& reactive powers calculated along the paths based on the provided lateral section loading values for the sections in which loads are present.
The equivalent impedance used while traversing across the feeder for radial paths (a combination of no-load nodes and nodes containing lateral feeders with loads) has been calculated using \eqref{Equiv_Impedance}.
\vspace*{-0.1cm}
\begin{equation}
z_{eq} = \sum\limits_{nl = 0}^{h_1} z_{n_l} + \frac{\sum\limits_{l_0 = 0}^{h_2} z_{l_0}}{h_2}
\label{Equiv_Impedance}
\end{equation}

where $z_{eq}$ is the equivalent impedance of the total radial path, $z_{nl}$  is the impedance in ohms in the $nl^{th}$ no load section, $z_{l_0}$ is the impedance in the $l_0^{th}$ section in which load is present, $h_1$ and $h_2$ are the number of no-load nodes and loaded nodes along the radial line respectively. It should be noted that the equivalent impedance for any parallel path in this topology is calculated as the mean of the equivalent impedances obtained from \textcolor{black}{the different paths constituting the parallel paths}.

\begin{figure}[h]
    \centering
    \includegraphics[width=0.5\textwidth]{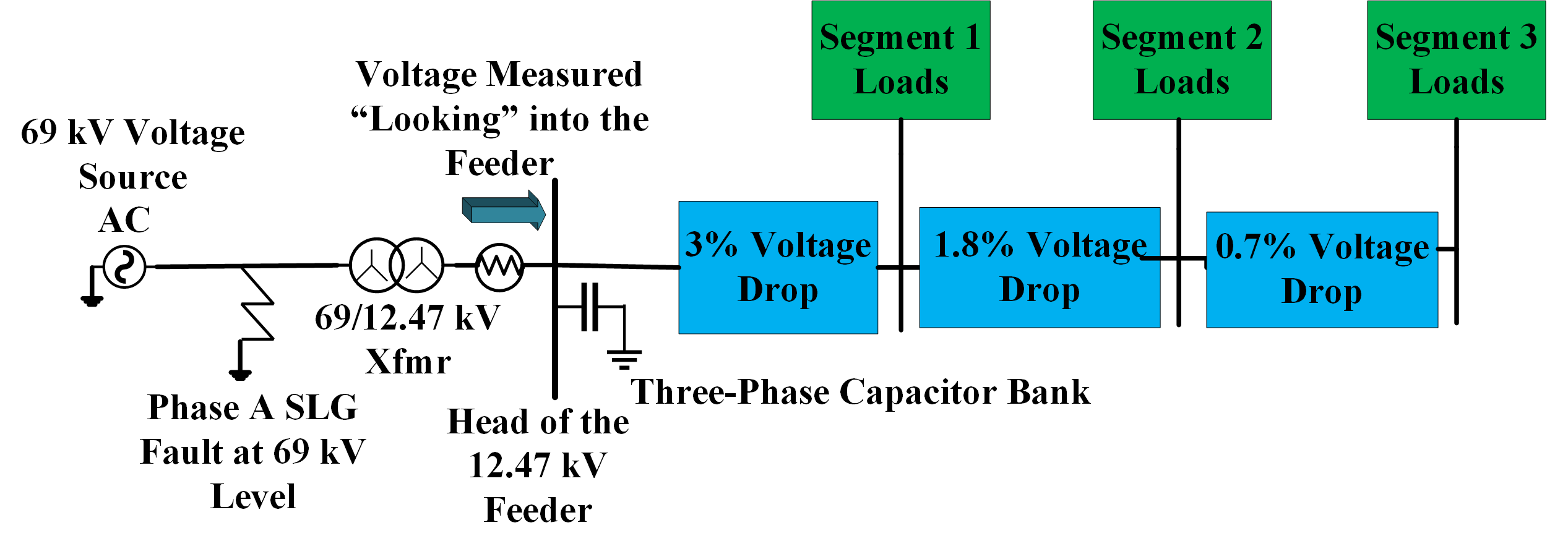}
    \caption{\textcolor{black}{Modified three-segment feeder model (feeder M) based on the local utility topological and loading data}} 
    \label{Fig3_Modified_3_segment_feeder_model_SRP}
     \vspace{-1em}
\end{figure}

The final three-segment feeder model (\textcolor{black}{denoted by} feeder M) obtained using the above steps is presented in Fig. \ref{Fig3_Modified_3_segment_feeder_model_SRP}. It should be noted that the Segment $k$ loads (where  $\; k \in \{1,2,3\}$)  correspond to the loads present in the $k^{th}$ segment of the feeder. \textcolor{black}{Additionally, the same} load types (SPHIM load, three-phase motor load, resistive load and the necessary distribution transformers to step down the voltage from the 12.47 kV level to its respective load voltage level) that are used in the original three-segment three-phase feeder model (\textcolor{black}{denoted by feeder O and as shown in Fig. \ref{Feeder_Structure}}) have been used here to construct \textcolor{black}{feeder M}. The voltage drops across the modified three-segment feeder model (feeder M) have been obtained using the proposed algorithm. It should be noted that in \textcolor{black}{feeder O} (in Fig. \ref{Feeder_Structure}), Segment 1 loads are placed at the head of the feeder and Segment 2, Segement 3 loads are placed across the two voltage drops (Voltage drop 1 and Voltage drop 2 in Fig. \ref{Feeder_Structure} correspond to 2.5\% and 1.3\% respectively) of the feeder. \textit{This clearly shows that the placement of the loads across the feeder M is significantly different using the proposed approach compared to the corresponding simplistic assumption made in the feeder O about the distribution of loads without validating it using utility topological and loading data.} 

Additionally, the final load composition obtained, for feeder M, using the proposed approach (in Fig. \ref{Fig3_Modified_3_segment_feeder_model_SRP}) is observed to be distributed as follows - Segment 1 (19\%), Segment 2 (35\%) and Segment 3 (46\%). From \cite{nekkalapu2022emtp} and \cite{nekkalapu2023development_etal}, the load composition distributed across the three segments in feeder O was assumed to be Segment 1 (30\%), Segment 2 (35\%) and Segment 3 (35\%) respectively \textcolor{black}{which is significantly different from the load composition obtained for feeder M}.   

\section{Simulations, Results \& Discussion }
\label{Section_Simu_and_Results}
Once the complete modified three-phase three-segment feeder model has been obtained using the proposed methodology, it is important to check the validity of the critical parameters for the contactor behavior and the SPHIM stalling parameters as defined in Section \ref{Section_contactor_modeling_background} which are obtained for Feeder O. For this reason, three scenarios each has been simulated for Feeder O and Feeder M by considering a SLG fault at 69 kV level on the sub-transmission side of the system as shown in Fig. 4 setup by using the same fault parameters - fault duration, fault impedance and fault angle (which corresponds to the location on the sine wave at which fault is initiated) for corresponding scenarios used for feeder O and feeder M respectively. The three scenarios simulated correspond to - Scenario 1 (severe fault which trips the contactors), Scenario 2 (moderately severe fault which makes the contactors chatter) and Scenario 3 (mild fault which has no impact on contactors).

\begin{table}[h]
\centering
\caption{Contactor and SPHIM stalling behavior comparison for feeder O \& feeder M}
\setlength{\tabcolsep}{3pt} 
\begin{tabular}{|c|c|c|c|c|c| c|c|} 
\hline
\begin{tabular}[c]{@{}l@{}}\textbf{Feeder Type/} \\ \textbf{ Scenario}  \end{tabular}  & \textit{\textbf{ST }}       & \begin{tabular}[c]{@{}l@{}}\textit{\textbf{T1}} \\ \textbf{(ms)}  \end{tabular} &  \begin{tabular}[c]{@{}l@{}}\textit{\textbf{V1}} \\ \textbf{(pu)}  \end{tabular} &  \begin{tabular}[c]{@{}l@{}}\textit{\textbf{T2}} \\ \textbf{(ms)}  \end{tabular} & \begin{tabular}[c]{@{}l@{}}\textit{\textbf{V2}} \\ \textbf{(pu)}  \end{tabular} & \centering \textit{\textbf{TMS}} & \textit{\textbf{IMS}} \\ \hline
O/1                    & Trips     &   11.6 & 0.89   & 79.8   &  0.89  &  3   & \{1,1,1\}    \\ \hline
O/2                    & Chatters  &   18.7 & 0.88   & 74.5   &   0.88 &  0   & \{0,0,0\}    \\ \hline
O/3                    & No affect &   - & -   & -   &  -  &  0   & \{0,0,0\}    \\ \hline
M/1                    & Trips     &   12.1 & 0.89   & 79.5   &  0.89  &  3   & \{1,1,1\}    \\ \hline
M/2                    & Chatters  &  31.6  & 0.86   & 84.3   &  0.90  &  2   & \{1,1,0\}   \\ \hline
M/3                   & No affect &   - & -   &  -  &  -  &  0  & \{0,0,0\}   \\ \hline
\end{tabular}
\label{Results}
\end{table}

From Table \ref{Results}, it can be observed that the contactor features and the motor stalling behavior is either same or very close for both the feeders for both Scenario 1 and Scenario 3. 
However, for Scenario 2, there is a significant difference between the observed trip and reconnection characteristics of the contactors between the feeders. This is to be expected because the voltage drops seen at the load terminals and the contactor terminals in each respective segment is different due to the different topology of the two feeders and the varied voltage drops along both the feeders. This clearly shows that the regression models proposed in \cite{nekkalapu2023development_etal} to develop a positive sequence contactor using feeder O would need to be retrained using feeder M. 

\begin{figure}[h]
    \vspace*{-1.1\baselineskip}
    \centering
    \includegraphics[width=0.42\textwidth]{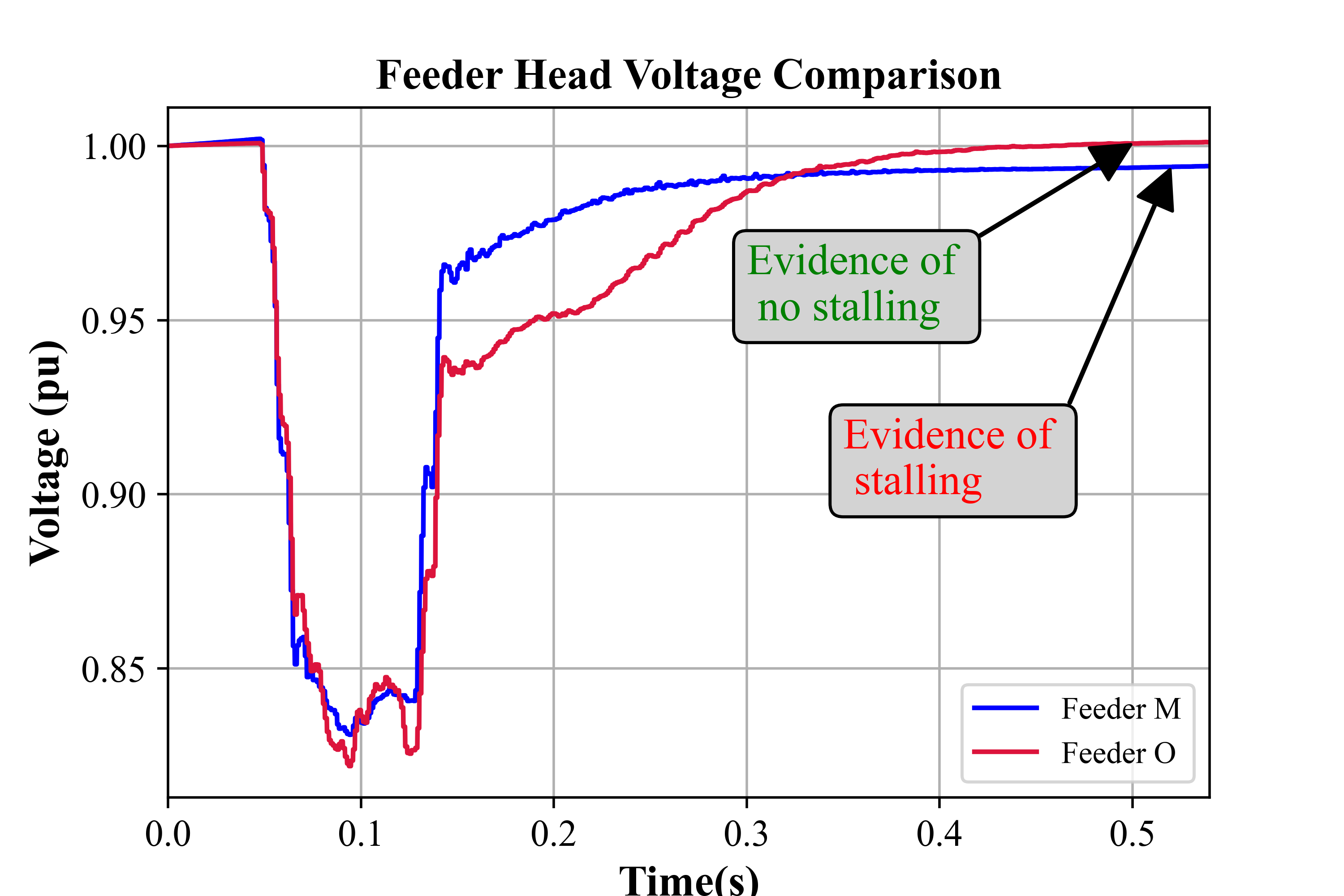} 
    \caption{\textcolor{black}{Comparison of positive sequence feeder voltages at feeder O \& feeder M for Scenario 2}} 
    \label{Feeder_Head_Voltage_Comparison}
    \vspace*{-1.5\baselineskip}
\end{figure}

\begin{figure}[h]
    \centering
    \includegraphics[width=0.42\textwidth]{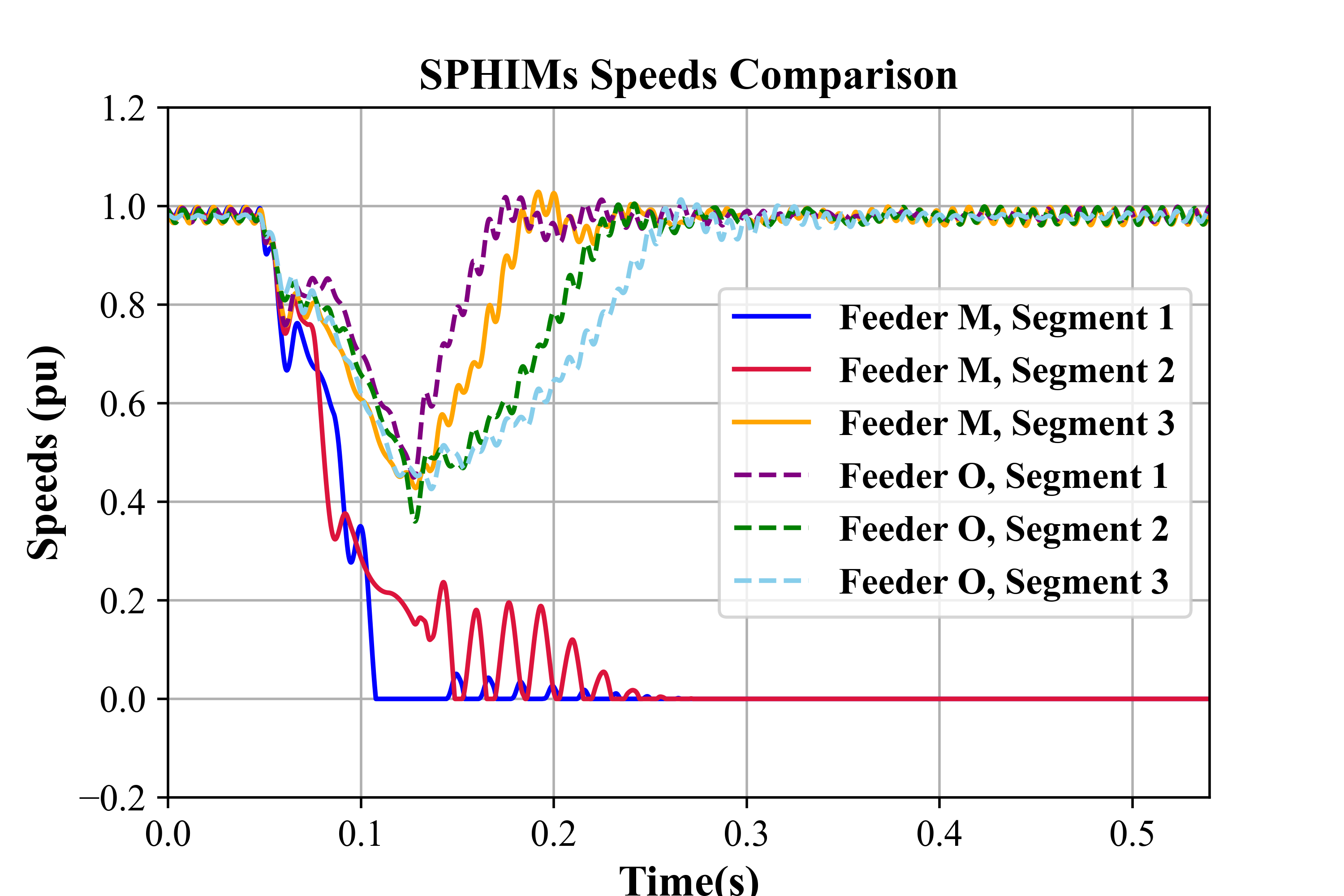} 
    \caption{\textcolor{black}{Evidence of SPHIM stalling in feeder M for Scenario 2}} 
    \label{SPHIM_Stalling}
     \vspace*{-1\baselineskip}
\end{figure}

In Scenario 2, from Fig. \ref{Feeder_Head_Voltage_Comparison}, it can be clearly seen that the positive sequence feeder head voltage for feeder O recovers back to its nominal voltage after the fault is cleared but the steady-state post-fault positive sequence feeder head voltage for feeder M is significantly lower compared to its corresponding feeder O voltage which clearly indicates SPHIMs in feeder M stalled leading to a FIDVR type critical phenomenon being experienced by the system. \textit{This clearly shows that the feeder O would have overestimated the feeder response for this fault event.} The evidence of SPHIMs being stalled in feeder M but not being stalled in feeder O has been demonstrated using Fig. \ref{SPHIM_Stalling}. In Fig. \ref{SPHIM_Stalling}, it can be seen that SPHIMs in segment 1 \& 2 stall for feeder M whereas all the SPHIMs recover back to their nominal speeds in feeder O after the fault is cleared in the system.

\section{Conclusions}
\label{Section_conclusions}
This paper develops a methodology to utilize the distribution feeders topological and loading data that are available to utilities to obtain a state-of-the-art reduced order three-segment feeder models. The utility of the proposed algorithm has been demonstrated by comparing the performance of the modified three-segment feeder model with the typically used three-segment feeder model in the literature, and it was observed that the latter model overestimates the feeder performance for moderately severe fault cases whereas the proposed feeder model demonstrates the ability to capture the SPHIMs stalling behavior accurately in cases where contactors usually chatters. This method enables utilities to generate an accurate representation of feeders for the analysis of large scale power systems without compromising on computational complexity issues that arises with representing the whole realistic feeder topology in system studies. Additionally, the developed algorithm is generic in nature and can also be utilized in planning studies to represent positive sequence feeder and load models.
\bibliographystyle{IEEEtran}
\bibliography{bibilography1}

\end{document}